\documentclass[12pt]{article}

\usepackage{axodraw}
\usepackage{amsmath}

\begin{document}
\baselineskip=20pt

\pagenumbering{arabic}

\vspace{1.0cm}
\begin{flushright}
LU-ITP 2004/043
\end{flushright}

\begin{center}
{\Large\sf Violation of Remaining Lorentz Symmetry in  the Approach of Time-Ordered Perturbation Theory to Space-Time Noncommutativity}
\\[10pt]
\vspace{.5 cm}

{Tobias Reichenbach\footnote{\emph{Email address:} tobias.reichenbach@itp.uni-leipzig.de}}
\vspace{1.0ex}

{\small Institut f\"ur Theoretische Physik, Universit\"at Leipzig,
\\
Augustusplatz 10/11, D-04109 Leipzig, Germany\\}

\vspace{2.0ex}

\end{center}

\begin{abstract}
We study remaining Lorentz symmetry, i.e. Lorentz transformations which leave the noncommutativity parameter $\theta^{\mu\nu}$ invariant, within the approach of time-ordered perturbation theory (TOPT) to space-time noncommutative theories. Their violation is shown in a simple scattering process. We argue that this results from the noncovariant transformation properties of the phase factors appearing in TOPT.
\end{abstract}

\begin{flushleft}


\end{flushleft}

\section{Introduction}
Noncommutative quantum field theory (NCQFT) has recently received renewed attention (see \cite{DouglasNekrasov} for a review). This interest is triggered by its appearance in the context of string theory \cite{SeibergWitten}, and by the observation that Heisenberg's uncertainty principle along with general relativity suggests the introduction of noncommutative space-time \cite{DFR}.

Coordinates are there considered as noncommuting Hermitian operators $\hat{x}^\mu$, which satisfy the commutation relation
\begin{equation}
[\hat{x}^\mu,\hat{x}^\nu]=i\theta^{\mu\nu}\quad.
\end{equation}
We will assume the antisymmetric matrix $\theta^{\mu\nu}$ to be constant. The algebra of these noncommuting coordinate operators can be realized on functions on the ordinary Minkowski space by introducing the Moyal $\star$-product
\begin{equation}
(f\star g)(x)=e^{\frac{i}{2}\theta^{\mu\nu}\partial_\mu^\xi\partial_\nu^\eta}f(x+\xi)g(x+\eta)\Big|_{\xi=\eta=0}
\quad .
\end{equation}
To obtain a NCQFT from a commutatitve QFT, one replaces the ordinary product of field operators by the star product in the action. Due to the trace property of the star product, meaning that
\begin{equation}
\int dx ~(f_1\star ... \star f_n)(x)
\end{equation}
is invariant under cyclic permutations, the free theory is not affected and noncommutativity only appears in the interaction part. As an example, the interaction in noncommutative $\varphi_\star^3$-theory reads
\begin{equation}
S_{\text{int}}=\frac{g}{3!}\int dx~(\varphi\star\varphi\star\varphi)(x)\quad . \\
\end{equation}

A first suggestion for perturbation theory has been made in \cite{Filk}, where the Feynman rules for the ordinary QFT are only modified by the appearance of momentum-dependent phase factors at the vertices. These are of the form
 $e^{-ip\wedge q}$, with $p\wedge q=\frac{1}{2}p_\mu\theta^{\mu\nu}q_\nu$.
Problems arise due to the nonlocality of the star product, which involves derivatives to arbitrary high orders. The S-matrix is no longer unitary in the case of space-time noncommutativity, i.e. $\theta^{0i}\neq 0$, as the cutting rules are violated \cite{GomisMehen}. \\

To cure this problem, a different perturbative approach, TOPT, has been suggested for scalar theories in \cite{LiaoSiboldTOPT}. It mainly builds on the observation that for space-time noncommutativity time-ordering and star product of operators are not interchangeable, their order matters. Defining TOPT by carrying out time-ordering \emph{after} taking star products, a manifestly unitary theory is obtained. The Feynman rules are considerably more complicated, ordinary propagators are no longer found but split up into two contributions. Another characteristic is the form of the phase factors, they depend on the internal momenta $q$ only through the on-shell quantities $q^\lambda=(\lambda E_q, \mathbf{q}), \lambda=\pm 1$.\\

However, further problems arise. In \cite{ORZ} it has been shown that Ward identities in NCQED are violated if TOPT is applied, which could be traced back to altered current conservation laws on the quantized level \cite{ReiDiplomarbeit}. In this paper, we want to prove another failure, the violation of remaining Lorentz symmetry in TOPT. \\

Space-time noncommutativity, meaning that time does not commute with space,  splits up into two cases, the so-called time-like and light-like one (see e.g. the discussion in \cite{HeslopSibold}). Here we consider the time-like one. In the standard form for this case $\theta^{\mu\nu}$ reads
\begin{equation}
\theta^{\mu\nu}= \begin{pmatrix} 0&\theta_e&0&0 \\ -\theta_e&0&0&0 \\ 0&0&0&\theta_m \\ 0&0&-\theta_m&0 \end{pmatrix}
\label{SpecForm}
\end{equation}
which remains invariant under transformations out of $SO(1,1)\times SO(2)\subset \mathcal{L_+^\uparrow}$. Therefore,  $SO(1,1)\times SO(2)$ is expected to be a remaining symmetry group. However, we will show in the following that this symmetry is not respected by TOPT.

Section 2 proves this statement by calculating a tree-level scattering amplitude in scalar noncommutative $\varphi_\star^3$-theory in two different frames, being related two each other by a transformation out of the above symmetry group. The results will differ from each other. We will argue in section 3 that this failure results from the non-covariant transformations of the phase factors.

\section{The violation in a scattering process}

To demonstrate the violation of remaining Lorentz invariance, we calculate a scattering amplitude in two different frames related by a remaining Lorentz transformation, and show that the results do not coincide. \\

We choose a two by two scattering process in noncommutative $\varphi^3$ theory, i.e. $\mathcal{L}_\text{int}=\frac{g}{3!}\varphi\star\varphi\star\varphi$, on tree-level for incoming on-shell momenta $p_1, p_2$ and outgoing momenta again $p_1,p_2$.
\begin{figure}[h]
\begin{center}
\begin{picture}(300,100)(0,0)
\Vertex(-5,50){1}
\Vertex(35,50){1}
\ArrowLine(-20,75)(-5,50)
\ArrowLine(-20,25)(-5,50)
\ArrowLine(35,50)(50,72)
\ArrowLine(35,50)(50,25)
\ArrowLine(-5,50)(35,50)
\Text(-25,85)[]{$p_1$}
\Text(-25,15)[]{$p_2$}
\Text(55,85)[]{$p_1$}
\Text(55,15)[]{$p_2$}
\Text(15,40)[]{$q_s$}
\Vertex(130,50){1}
\Vertex(170,50){1}
\ArrowLine(115,75)(130,50)
\ArrowLine(115,25)(170,50)
\ArrowLine(170,50)(185,75)
\ArrowLine(130,50)(185,25)
\ArrowLine(130,50)(170,50)
\Text(110,85)[]{$p_1$}
\Text(110,15)[]{$p_2$}
\Text(190,85)[]{$p_1$}
\Text(190,15)[]{$p_2$}
\Text(150,60)[]{$q_u$}
\Vertex(285,65){1}
\Vertex(285,35){1}
\ArrowLine(250,75)(285,65)
\ArrowLine(250,25)(285,35)
\ArrowLine(285,65)(320,75)
\ArrowLine(285,35)(320,25)
\ArrowLine(285,65)(285,35)
\Text(245,85)[]{$p_1$}
\Text(245,15)[]{$p_2$}
\Text(325,85)[]{$p_1$}
\Text(325,15)[]{$p_2$}
\Text(295,50)[]{$q_t$}
\end{picture}
\parbox{12cm}{\caption{\label{scatt}  A scattering process in $\varphi^3$ theory: s-, u- and t-channel}}
\end{center}
\end{figure}
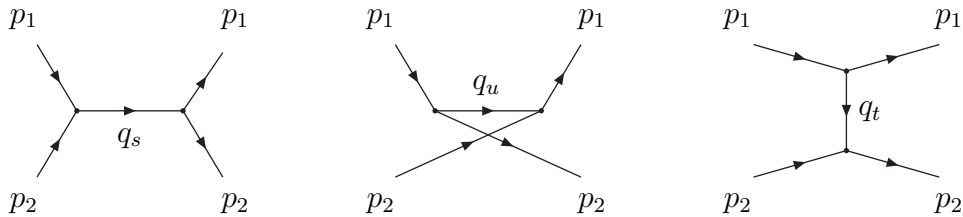
The amplitude is diagrammatically given by the graphs in Fig. \ref{scatt}, and according to TOPT (see \cite{LiaoSiboldTOPT} for details) corresponds to the analytic expressions
\begin{align}
i\mathcal{M}&=i\mathcal{M}_s+i\mathcal{M}_u+i\mathcal{M}_t \cr
i\mathcal{M}_s&=g^2\sum_{\lambda=\pm 1}\frac{1}{2E_{q_s}}\frac{\lambda}{q^0_s-\lambda(E_{q_s}-i\epsilon)}V(p_1,p_2,-q_s^\lambda)^2\Big|_{q_s=p_1+p_2}  \cr
i\mathcal{M}_u&=g^2\sum_{\lambda=\pm 1}\frac{1}{2E_{q_u}}\frac{\lambda}{q^0_u-\lambda(E_{q_u}-i\epsilon)}V(p_1,p_2,-q_u^\lambda)^2\Big|_{q_u=p_1-p_2} \cr
i\mathcal{M}_t&=g^2\sum_{\lambda=\pm 1}\frac{1}{2E_{q_t}}\frac{\lambda}{q^0_t-\lambda(E_{q_t}-i\epsilon)}V(p_1,p_2,-q_t^\lambda)^2\Big|_{q_t=0}
\label{iM}
\end{align}
where
\begin{align}
E_q&=\sqrt{m^2+\mathbf{q}^2} \cr
q^\lambda&=(\lambda E_q,\mathbf{q}) \cr
V(p_1,p_2,p_3)&= \frac{1}{6}\sum_{\pi\epsilon S_3} e^{-i(p_{\pi(1)}, p_{\pi(2)}, p_{\pi(3)})}\quad,
\end{align}
the phase factor $V$ is written with help of the abbreviation
\begin{align}
(p_1,...,p_n)=\sum_{i<j}p_i\wedge p_j\quad .
\end{align}

Now, we will choose specific $p_1, p_2$ and $\theta^{\mu\nu}$ of type (\ref{SpecForm}) in frame 1, calculate $i\mathcal{M}=i\mathcal{M}_s+i\mathcal{M}_u+i\mathcal{M}_t$ there and compare to $i\mathcal{M}'$ which we compute in frame 2 being related to frame 1 by the transformation
\begin{equation}
G=\begin{pmatrix} \cosh\beta&\sinh\beta&0&0 \\ \sinh\beta&\cosh\beta&0&0\\ 0&0&1&0\\ 0&0&0&1 \end{pmatrix}
\quad \in\quad SO(1,1)\times SO(2) \quad .
\label{transf}
\end{equation}

The following configuration is chosen in frame 1:
\begin{align}
p_1&=(E_p,0,0,p)\quad, \qquad E_p=\sqrt{m^2+p^2} \cr
p_2&=(E_p,0,0,-p) \cr
\theta^{\mu\nu}&=\theta_e\begin{pmatrix} 0&1&0&0 \\ -1&0&0&0 \\ 0&0&0&0 \\ 0&0&0&0 \end{pmatrix} \quad ,
\end{align}
such that the internal momenta are
\begin{align}
q_s&=p_1+p_2=(2E_p,0,0,0) \cr
q_s^\lambda&=(\lambda m,0,0,0) \cr
q_u&=p_1-p_2=(0,0,0,2p) \cr
q_u^\lambda&=(\lambda E_{2p},0,0,2p)\quad ,\qquad E_{2p}=\sqrt{m^2+(2p)^2} \cr
q_t&=p_1-p_1=(0,0,0,0) \cr
q_t^\lambda&=(\lambda m,0,0,0)  \quad .
\end{align}
We find the configuration in frame 2 by applying the transformation (\ref{transf}):
\begin{align}
p_1'&=(E_p\cosh\beta,E_p\sinh\beta,0,p)  \cr
p_2'&=(E_p\cosh\beta,E_p\sinh\beta,0,-p) \cr
\theta^{'\mu\nu}&=\theta^{\mu\nu}=\theta_e\begin{pmatrix} 0&1&0&0 \\ -1&0&0&0 \\ 0&0&0&0 \\ 0&0&0&0 \end{pmatrix} \quad,
\end{align}
implying the internal momenta
\begin{align}
q'_s&=p_1'+p_2'=(2E_p\cosh\beta,2E_p\sinh\beta,0,0) \cr
(q'_s)^\lambda &= (\lambda E_{q'_s}, 2E_p\sinh\beta,0,0)\quad, \quad E_{q'_s}=\sqrt{m^2+4E_p^2\sinh^2\beta} \cr
q'_u&=p_1'-p_2'=(0,0,0,2p) \cr
(q'_u)^\lambda&=(\lambda E_{2p},0,0,2p)\cr
q'_t&=p_1'-p_1'=(0,0,0,0) \cr
(q'_t)^\lambda&=(\lambda m,0,0,0)
\qquad .
\end{align}
In frame 1, we note that the phase factors relevant for $i\mathcal{M}_s, i\mathcal{M}_u$ and $i\mathcal{M}_t$ do not involve $\theta^{\mu\nu}$, such that the result is the same as in the commutative case:
\begin{align}
i\mathcal{M}_u&=g^2\frac{1}{(p_1+p_2)^2-m^2} \\
i\mathcal{M}_u&=g^2\frac{1}{(p_1-p_2)^2-m^2} \\
 i\mathcal{M}_t&=-g^2\frac{1}{m^2} \quad .
\end{align}
However, in frame 2 the phase factors do involve $\theta^{\mu\nu}$, we may expand the resulting amplitudes to second order in $\theta_e$ and arrive after some calculation (see \cite{ReiDiplomarbeit} for details) at
\begin{align}
i\mathcal{M'}_s
=~&g^2\frac{1}{(p_1+p_2)^2-m^2}
-\frac{2}{3}g^2\theta_e^2E_p^2\frac{m^2+4E_p^2\sinh^2\beta}{3m^2+4p^2}\sinh^2\beta +o(\theta_e^2)\\
i\mathcal{M'}_u
=~&g^2\frac{1}{(p_1-p_2)^2-m^2}+\frac{2}{3}g^2\theta_e^2 E_p^2\sinh^2\beta +o(\theta_e^2)\\
i\mathcal{M'}_t=~&-g^2\frac{1}{m^2}+\frac{2}{3}g^2\theta_e^2E_p^2 \sinh^2\beta +o(\theta_e^2)\quad .
\end{align}
The difference between the amplitudes in the two different frames can easily be read off, further manipulation yields
\begin{align}
i\mathcal{M}-i\mathcal{M}'
&=\frac{2}{3}g^2\theta_e^2 E_p^2\Big(\frac{m^2+4E_p^2\sinh^2\beta}{3m^2+4p^2}-2\Big)\sinh^2\beta+o(\theta_e^2)\\
&< 0 \qquad \quad \text{for}\quad p,\theta_e,\beta \neq 0 \quad \text{and} ~ \beta ~\text{sufficiently small.} \nonumber
\end{align}

The last inequality is easily verified if we notice that for small enough $\beta$ the term $\frac{m^2+4E_p^2\sinh^2\beta}{3m^2+4p^2}$ is less or equal to $\frac{1}{3}$. We have thus shown that the scattering amplitude differs in two frames which are related by a remaining symmetry transformation.

\section{Non-covariant transformation of the phase factors}

The origin of the above demonstrated violation of remaining symmetry in TOPT lies in the non-covariant transformation of the phase factors in TOPT.\\
 Let $p_i$ be the external, $q_j$ the internal momenta, the phase factors then depend on $p_i$ and $q_j^{\lambda_j}$, where $\lambda_j=\pm1$. More precisely, these phase factors are functions of the complex numbers $p_i\wedge p_j, p_i\wedge q_j^{\lambda_j}$ and $ q_i^{\lambda_i}\wedge q_j^{\lambda_j}$. Under a transformation that leaves $\theta^{\mu\nu}$ unchanged and takes $p_i \rightarrow p_i',~ q_j \rightarrow q_j'$ we have
\begin{eqnarray}
p_i\wedge p_j~ \rightarrow& p_i'\wedge p_j'&=~p_i\wedge p_j \cr
p_i\wedge q_j^{\lambda_j}~ \rightarrow &p_i'\wedge (q_j')^{\lambda_j}& \neq ~p_i'\wedge (q_j^{\lambda_j})'\hspace{6mm}=~p_i\wedge q_j^{\lambda_j} \cr
q_i^{\lambda_i}\wedge q_j^{\lambda_j}~ \rightarrow &(q_i')^{\lambda_i}\wedge (q_j')^{\lambda_j} &\neq ~(q_i^{\lambda_i})'\wedge (q_j^{\lambda_j})'=~q_i^{\lambda_i}\wedge q_j^{\lambda_j}
\end{eqnarray}
where the inequalities in the last two lines arise because the internal momenta $q_i$ are in general \emph{not} on-shell and therefore $(q_i')^{\lambda_i}\neq (q_i^{\lambda_i})'$. This means that the noncommutative phase factor is not left invariant by the transformation and can lead, as demonstrated above, to different amplitudes.

\section{Conclusions}

Although TOPT solves the unitarity problem in scalar space-time noncommutative theories, it poses further problems. At first, internal symmetries are altered, as it has been shown that Ward identities in gauge theories are not longer valid \cite{ORZ}. In this letter we have addressed another problem, the violation of remaining Lorentz symmetry. This result may  suggest to modify time-ordering in a way which explicitly preserves remaining Lorentz symmetry. Such work has been carried out recently \cite{HeslopSibold}, internal symmetries are also investigated in the new approach.

\section*{Acknowledgements}
I am grateful to K. Sibold, Y. Liao, P. Heslop and C. Dehne for fruitful discussions and conversations that stimulated this work.

\end{document}